# Ultrafast Optical Signal Processing with Bragg Structures

**Yikun Liu[1], Shenhe Fu[2], Boris A. Malomed[3,4], Iam Choon Khoo[5] andJianying Zhou[1,*]**

[1] State Key Laboratory of Optoelectronic Materials and Technologies, Sun Yat-sen University, Guangzhou 510275, China; liuyk6@mail.sysu.edu.cn
[2] Department of Optoelectronic Engineering, Jinan University, Guangzhou 510632, China; fushenhe@jnu.edu.cn
[3] Department of Physical Electronics, School of Electrical Engineering, Faculty of Engineering, Tel Aviv University, Tel Aviv 69978, Israel; malomed@post.tau.ac.il
[4] Laboratory of Nonlinear-Optical Informatics, ITMO University, St. Petersburg 197101, Russia
[5] Electrical Engineering Department, Pennsylvania State University, University Park, PA 168 , USA; ick1@psu.edu
[*] Correspondence: stszjy@mail.sysu.edu.cn; Tel.: +86-20-8411-0277

**Abstract:** The phase, amplitude, speed, and polarization, in addition to many other properties of light, can be modulated by photonic Bragg structures. In conjunction with nonlinearity and quantum effects, a variety of ensuing micro- or nano-photonic applications can be realized. This paper reviews various optical phenomena in several exemplary 1D Bragg gratings. Important examples are resonantly absorbing photonic structures, chirped Bragg grating, and cholesteric liquid crystals; their unique operation capabilities and key issues are considered in detail. These Bragg structures are expected to be used in wide-spread applications involving light field modulations, especially in the rapidly advancing field of ultrafast optical signal processing.

**Keywords:** Bragg structure; ultrafast; optical signal process

## 1. Introduction

In the next generation of high-speed information networks, the direct processing of optical signals is required. On the other hand, the basic signal-processing capabilities, such as switching, logic operations, and buffering, are still lacking in practically usable forms. Useful for achieving these objectives should be the deceleration of optical signals and the creation of standing ones. These effects have been demonstrated with the help of various techniques, such as electromagnetically-induced transparency [1,2], but those interference-based techniques are often not suitable for broadband signal processing, when the carrier waves are represented by picosecond or even femtosecond pulses [2]. The light can also be retarded in optical fibers by a stimulated Brillouin scattering effect [3], and in photonic-crystal waveguides by manipulating the dispersion [4,5]. Light-matter interactions can be enhanced by the retardation of light. Since the light-matter interaction time $t$ is inversely proportional to the group velocity $v_g$, the use of slow light with a small $v_g$ implies longer interaction times, and consequently, a more efficient energy conversion [6]. Slow light also offers the possibility to compress optical signals in space, thus reducing the device size [7,8].

In this article, we review both theoretical and experimental results concerning the processing of ultrafast optical signals in one-dimensional (1D) Bragg gratings, which exhibit a 1-D photonic bandgap that makes it possible to significantly reduce the speed of light launched at a carrier frequency close to the bandgap. We discuss both artificially engineered Bragg structures, made of optoelectronic materials, and those produced by natural self-assembly, such as cholesteric liquid crystals. We provide a critical review of the performance of the Bragg gratings and the limitations in their use. In particular, we describe how some of the inevitable deleterious effects that accompany a strong dispersion experienced by light at the band-edge, can be balanced by the material nonlinearity, which provides the laser-induced self-phase modulation of the optical field. Theoretical modeling of these processes in some Bragg structures have shown that ultrafast laser pulses can be decelerated, stopped, and buffered; as a result, stationary nonlinear optical frequency conversion can be very efficient, even with a very thin resonant absorption Bragg reflector sample [9–13]. In a tailored optical structure, such as one into which defects and spatial chirp are integrated, various optical logic operations can be efficiently realized [14–16]. Recently, BG structures consisting of cholesteric liquid crystals (CLCs), which possess extraordinarily large ultrafast optical nonlinearities due to photonic crystal band-edge enhancement, have also been shown to be highly effective for direct-action compression, or for the stretch and recompression of pico- and femto-second pulses, opening up new possibilities for efficient broadband optical-signal processing [17,18].

## 2. The Bragg-Grating (BG) Structure

Bragg structures are basically 1D photonic crystals (PCs) [19,20] comprising materials with periodic refract index modulation along one direction, taken as:

$$n(z) = n_0 \left[1 + a_1 \cos(2k_c z)\right], \tag{1}$$

with constant $k_c$, $a_1$, and $n_0$ values. The fabrication, characterization, and optical properties of PCs have been thoroughly investigated since the original works of John and Yablonovitch [19,20]. It is well known that the BG structure gives rise to a bandgap, with its central wavelength located at $\lambda_0 = \dfrac{2\pi n_0}{k_c}$ and a bandwidth of $\Delta\lambda = \dfrac{2\pi n_0 a_1}{k_c}$. Strong dispersion and velocity reduction occur at the photonic band edges [21]; in some materials, losses can also be significant. Both a low group velocity and low dispersion can be obtained by designing the structures of BG. In Reference [22], a low velocity of 0.02*c* (*c* is light speed in vacuum), in combination with a 10 nm-bandwidth and low dispersion, was demonstrated by changing the structure of a PC's waveguide [22].

A frequently studied 1-D Bragg structure is the fiber Bragg grating, which can be fabricated by interference lithography in fibers [23]. Fiber Bragg gratings are widely used in sensors [24], optical telecommunications [25], and for dispersion compensation [26]. 1D Bragg structures can also be fabricated in other solid materials, such as silicon [27] and AlGaAs [28], for optical switching and limiting operations [29]. Besides such artificial Bragg structures, self-assembled Bragg structures can also be found in liquid-crystal materials, such as cholesteric liquid crystals, whose optical properties can be modulated by the light field [30]. Cholesteric liquid crystals can also be used as temperature sensors [31], due to their temperature-dependent pitch.

By introducing nonlinear optical effects, many interesting discoveries have been made in one-dimension (1D) fiber Bragg grating. These include multistability, a zero velocity, and the creation of wobbling or oscillating solitons [32–36]. Using Kerr nonlinearity to balance the strong dispersion near the bandgap's edge, BG solitons with speeds of 0.5 *c* [35,37] and as low as *c*/7 [33] were observed in experiments. Furthermore, it was predicted that standing light can be created using BG fibers with defects [38–40], Bragg reflectors combined with resonant nonlinearity [41], and the collision of BG solitons [42][43]. However, high input power densities (>10 GW/cm$^2$) and long propagation lengths are required to achieve strong nonlinear effects, which may pose serious problems in experiments and applications.

Several Bragg structures have been evoked to address the power and interaction length issues, as detailed in the following sections.

## 3. Optical Signal Processing in Resonantly Absorbing Bragg Reflector (RABR)

*3.1. Theoretical Considerations*

RABR is produced by adding narrow stripes, doped by two-level atoms resonantly interacting with the transmitted electromagnetic fields, to a BG structure, as schematically shown in Figure 1 [9–11].

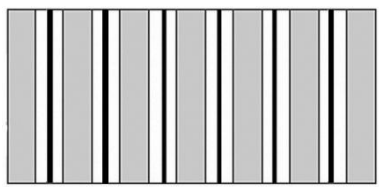

**Figure 1.** A scheme of RABR with black stripes representing thin layers of two-level atoms. White and gray bands represent a periodically structured nonabsorbing medium (After Reference [11]).

Light-matter interaction in RABR, built of infinitely thin atomic layers, is modeled by the Maxwell-Bloch equations for electric-field components $E^+$ and $E^-$ of the forward- and backward-propagating waves, and population W [12]:

$$\frac{\partial \Sigma^\pm}{\partial \tau} \pm \frac{\partial \Sigma^\pm}{\partial \zeta} = i\eta\Sigma^\pm + P, \qquad ()(2)$$

$$\frac{\partial P}{\partial \tau} = -i\delta P + (\Sigma^+ + \Sigma^-)W \qquad (3)$$

$$\frac{\partial W}{\partial \tau} = -\mathrm{Re}\left[(\Sigma^+ + \Sigma^-)P^*\right] \qquad (4)$$

where $\Sigma^\pm \equiv (2\tau_c \mu/\hbar)E^\pm$, $\tau_c \equiv (2\hbar n_0 / \mu_0 c^2 \omega_c \mu \rho)^{1/2}$ is the cooperative time (ranging from pico- to femto-seconds) which is determined by the presence of the two-level atoms; $n_0$ is the average refraction index; $\omega_c$ is the resonant frequency of the two-level atom; μ is the magnitude of the dipole matrix element ρ is the density of the dopant atoms; $\eta \equiv (n_1 \omega \tau_c)/4$ is the dimensionless coupling constant; *P* and *W* are the material polarization and population inversion density, respectively; $\delta \equiv (\omega - \omega_c)/\tau_c$ is the dimensionless detuning; $\tau \equiv t/\tau_c$ and $\zeta \equiv (n_0/c\tau_c)z$ are the dimensionless time and spatial coordinates, respectively; and ω is the frequency of the incident light. Here, $\tau_c$ is equal to 300 fs [13].

Solutions of the Maxwell-Bloch equations, some being available in an analytical form [10], produce a vast family of stable gap solitons, of both a standing and moving nature. Compared to the self-induced-transparency (SIT) in uniform media, the solitons generated in RABR may have an arbitrary pulse area, while in uniform media solitons, are only created by pulses with an area exactly equal to 2π [44,45]. Stable dark solitons can also be excited in RABR [11]. Thus, the theoretical analysis demonstrates that light signals with any velocity can indeed exist in the RABR.

An optical pulse with a hyperbolic-secant shape, generated by the input with a small area, undergoes complete Bragg reflection in the RABR. With an increase in the input intensity, the SIT solitons can be excited, making it possible for light pulses to propagate without loss. If the intensity is still higher, the splitting of the pulse occurs. Most remarkably, with a suitable incident pulse, an oscillating gap soliton trapped in the RABR as a standing wave can be created. Figure 2a shows the evolution of the optical energy, which is represent by the population distribution *W* [12,13], in this case. Furthermore, we have found that multiple gap solitons can be simultaneously created as standing modes, thus predicting a possibility of efficient storage of the optical energy in the RABR structure (Figure 2b).

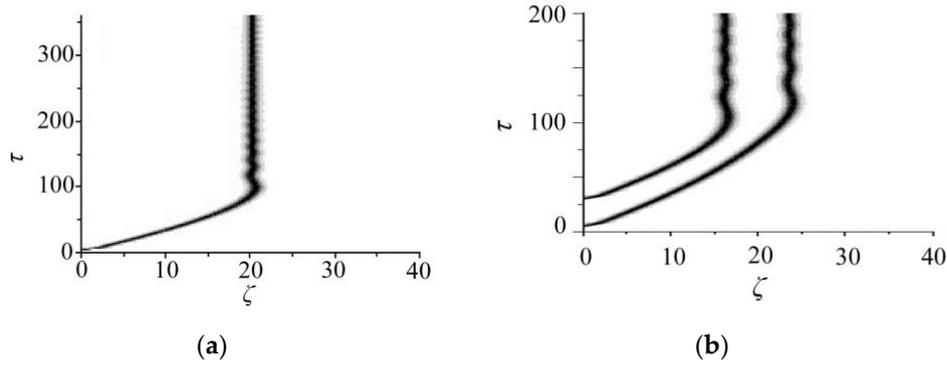

(a)                                (b)

**Figure 2.** (**a**) The evolution of the population inversion, *W*, illustrating the creation of an oscillatory standing soliton in the RABR from a sech-shaped input. The pulse width is $\tau_0 = 0.5\tau_c$, and the injected amplitude is $\Sigma_0^+ = 4.3$; (**b**) The generation of two standing oscillatory solitons, with the pulse widths $\tau_0 = 0.5\tau_c$ and injected amplitude $\Sigma_0^+ = 3.6$ [12].

For an input pulse with a suitable area, the numerical simulations have shown that the pulse can evolve into a standing gap soliton that can be stored as a stable self-localized state. The minimum length of the sample, necessary for the realization of the storage of the standing light pulse, is 1200 BG periods [12,13]. The self-trapping dynamics of the laser pulse may be considered as the motion of a quasi-particle in an effective potential representing the nonlinear interaction between the input pulse and the two-level atomic medium [46]. As a result, the pulse decelerates and eventually comes to a halt inside the structure, under the action of the trapping potential.

By colliding with another input pulse, the stored pulse can be released; both pulses can be released from the RABR with an efficiency of up to 96% (Figure 3) [13].

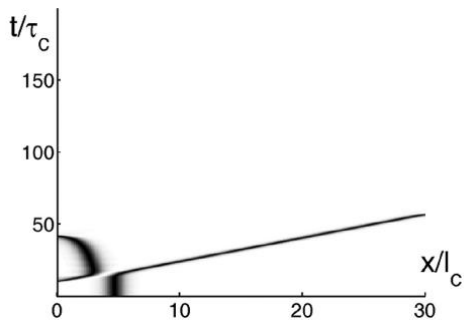

**Figure 3.** A contour plot illustrating the release of a laser pulse stored in the RABR by dint of its collision with an additional pulse injected into the structure [13]. The pulse width is $\tau_0 = 0.5\tau_c$, and the injected amplitude is $\Sigma_0^+ = 3.5$, for the stopped pulse, and $\Sigma_0^+ = 3.84$ for the incident one. Both input pulses have a standard sech profile.

The nonlinear frequency conversion of zero-velocity (standing) short light pulses, based on the stimulated Raman scattering (SRS), has also been studied [7]. To reduce the walk-off effects, both the pump and Stokes waves must be phase-matched, traveling with equal speeds. This can be realized in a 1D doubly resonant Bragg reflector (DRBR), where "doubly" implies supporting the resonant interactions at two different wavelengths simultaneously, as specified below [7].

The DRBR structure consists of 1D periodically arranged layers of two-level atoms and a passive BG. The period of the array of the two-level-atom layers is equals to the half-wavelength of the pump pulse. The bandgap center of the passive BG is equal to the wavelength of the Stokes pulse. In the DRBR, the pump pulse can be decelerated and stopped by thin atomic layers [41]. On the other hand, the Stokes pulse can be generated as a standing or slowly oscillating soliton by the passive Bragg reflector [14]. Due to the interaction of these two kinds of stopped light pulses, the energy of the pump pulse can be efficiently transferred to the Stokes pulse. Furthermore, the energy of the Stokes pulse will eventually leak out from both edges of the finite-length of DRBR in the form of Raman solitons.

The propagation of light in the DRBR is modeled by the Maxwell-Bloch equations, which take into account the Raman and Kerr nonlinearities [7]:

$$\pm \frac{\partial \Sigma_p^\pm}{\partial \zeta} + \frac{\partial \Sigma_p^\pm}{\partial \tau} = -\frac{G_p}{2}\Sigma_p^\pm (|\Sigma_s^+|^2 + |\Sigma_s^-|^2) + P + i\Gamma_p \Sigma_p^\pm \left[ |\Sigma_p^\pm|^2 + |\Sigma_p^\mp|^2 + (2-f_R)(|\Sigma_s^+|^2 + |\Sigma_s^-|^2) \right] \quad (5);$$

$$\pm \frac{\partial \Sigma_s^\pm}{\partial \zeta} + \frac{\partial \Sigma_s^\pm}{\partial \tau} = -\frac{G_s}{2}\Sigma_s^\pm (|\Sigma_p^+|^2 + |\Sigma_p^-|^2) + iK\Sigma_s^\mp + i\Delta\Sigma_s^\pm + i\Gamma_s \Sigma_s^\pm \left[ |\Sigma_s^\pm|^2 + |\Sigma_s^\mp|^2 + (2-f_R)(|\Sigma_p^+|^2 + |\Sigma_p^-|^2) \right] \quad (6);$$

$$\frac{\partial P}{\partial \tau} = (\Sigma_p^+ + \Sigma_p^+)W, \quad (7);;$$

$$\frac{\partial W}{\partial \tau} = -\mathrm{Re}\left[(\Sigma_p^+ + \Sigma_p^+)P^*\right], \quad (8);$$

where $\Sigma_{p,s}^\pm \equiv (2\tau_c \mu/\hbar)E_{p,s}^\pm$ represents the forward- and backward-propagating pump and Stokes waves; $K \equiv \kappa c\tau_c/n_0$ and $\Delta \equiv \delta c\tau_c/n_0$ are the dimensionless coupling constants and detuning, respectively; $f_R$ is the fraction of the nonlinearity arising from molecular vibrations with a typical value of 0.18 [45]; and $G_{p,s} \equiv \left(\varepsilon_0 c^2 \hbar^2 / 8\mu^2 \tau_c\right)g_{p,s}$ and $\Gamma_{p,s} \equiv \omega_{p,s} n_2 \varepsilon_0 c\hbar^2 / 8\mu^2 \tau_c$ are the dimensionless nonlinearity and gain coefficients for the pump and Stokes waves, respectively.

When the time of the interaction of the standing Raman active medium and soliton is large enough, the intensity of the zero-velocity Stokes pulse starts to increase at the expense of the pump field. Following the efficient power exchange between the pump and Stokes pulses, the energy of the Stokes pulse can leak out from both edges of the DRBR. By comparing the output energy of the Stokes pulses with the energy of the standing pump pulse, the efficiency of the Raman shift can be estimated (Figure 4). Remarkably, it may exceed 85%, i.e., higher than the efficiency in the bulk medium [47]. In practice, such an efficient conversion can be achieved by using a periodically arranged multi-quantum-well structure, [48]. The required length and power are only a few millimeters and 100 μJ, respectively.

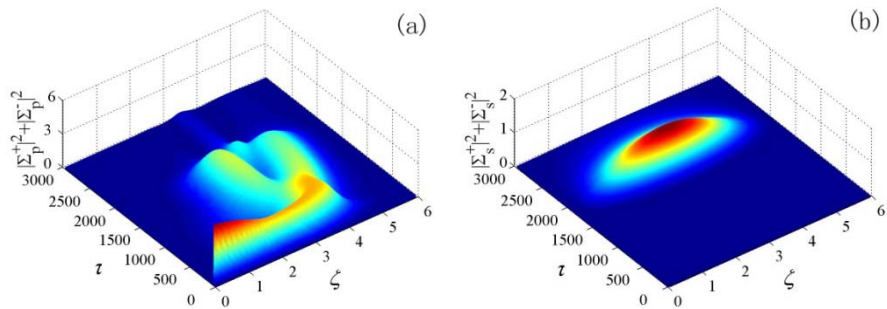

**Figure 4.** The evolution of the energy density of (**a**) pump and (**b**) stokes pulses in the DRBR under the action of the SRS [7].

RABR can also be used to reshape and compress ultrashort laser pulses. From the Maxwell-Bloch equations (1), one can show that the RABR may compress the sech-shaped input, with a pulse width ranging from $2\tau_c$ to $5\tau_c$ ($\tau_c$ is the cooperative time), into a $2\pi$ SIT soliton [49]. Furthermore, input pulses with a multiple-peak shape can be re-shaped to produce a single-peak output. Figure 5 shows the transformation of the three-peak pulse into a single-peak pulse in the RABR structure. In fact, SIT in bulk media can also be applied to compressed optical pulses. However, in that case, the input pulse with single peak is split into multiple ones, rather than morphed into a single-peak SIT soliton, as in RABR. The difference is explained by the fact that reshaping in the RABR originates from multiple reflections in the Bragg structure, while no reflections occur in the the bulk SIT medium. Other interesting phenomena, such as the negatively refracted light in the RABR, have also been discovered [50].

The validity of the numerical predictions was also checked by the comparison with a more realistic finite-atomic-width approach [51]. The result shows that stable moving solitons can be generated, but the zero-velocity soliton no longer exists if the thickness of the two-level atom layer exceeds a critical value (1.2 nm).

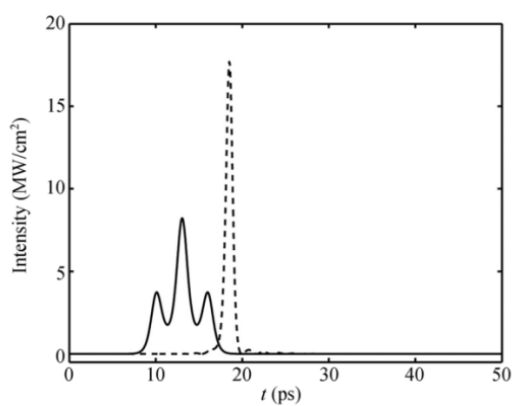

**Figure 5.** The input (solid line) and output (dash line) from the RABR, when a three-peak pulse is used as the input[40].

*3.2. Experimental Work on RABR*

Many methods, such as photo lithography, e-beam, and self-assembling, can be used to fabricate 1D photonic structures. Due to the extremely thin layers filled by dopant atoms, the molecular-beam epitaxy technique is particularly suitable for fabricating RABRs. In Reference [51], InGaAs/GaAs multi-quantum-well structures, which may be seen as typical RABRs, have been made by using this technique. In a 200 layer InGaAs/GaAs multi-quantum-well pattern, ultrafast switching based on the ac Stark has been observed[52].

For the purpose of buffering the optical pulse, the number of layers should be increased to about 1000 [43]. In this case, the sample will be very easy to peel off from the substrate with the help of demoulding. Normally, for measuring the nonlinear effect of the InGaAs/GaAs multi-quantum-well, the experiment should be performed at a low temperature (4K–10 K) [48,52]. These strict experimental conditions limit the application of InGaAs/GaAs multi-quantum-wells in optical signal processing.

**4. Optical Signal Processing in Chirped Bragg Structures**

Chirped fiber BGs with a gradually varying local BG period, have been widely used to provide a strong positive/negative dispersion in a fiber system to stretch and compress the optical pulse [26]. Recently, the nonlinear propagation of the optical pulse in a silicon chirp Bragg structure was investigated[14,15]. The simulation result shows that the optical pulse can be buffered and released in the silicon chirp Bragg structure.

Generally, for the purpose of creating very slow BG solitons, two conditions should be met. First, the grating-induced dispersion must be balanced by the nonlinearity [53]. Second, since the optical field in the 1D Bragg structure is represented by forward-traveling and backward-traveling waves, an initial configuration should have nearly-equal powers in the two waves [54]. The former condition can be achieved by using a high-power input, and therefore, sufficient nonlinearity [12]. The latter condition is much harder to meet with the single incident pulse [55], as it does not initially contain any backward component.

In other studies [14,15], concatenated BGs were used to generate slow ultrashort pulses, or even standing ones. Such concatenated BGs are built by linking a linearly chirped grating with a uniform one. The main advantage of this setting is that the initial conditions for pulses at the input edge of the uniform BG segment may be manipulated by means of the preceding chirped BG, which provides a possibility of preparing the right mix of forward and backward fields.

As in Bragg superstructures, light propagation in chirped gratings can be described by the standard coupled-mode theory [53,56–59]. For slowly varying envelopes of forward and backward waves, $E_f$ and $E_b$, the coupled-mode equations are written as [14,56]:

$$\pm i \frac{\partial E_{f,b}}{\partial z} + \frac{i}{v_g} \frac{\partial E_{f,b}}{\partial t} + \delta(z) E_{f,b} + \kappa(z) E_{b,f} + \gamma (|E_{f,b}|^2 + 2|E_{b,f}|^2) E_{f,b} = 0, \qquad (9)$$

where $t$ is the time; $v_g = c/n_0$ is the group velocity in the material of which the BG is fabricated; and $\gamma = n_2 \omega / c$ is the nonlinearity strength, where ω is the frequency of the carrier wave and $n_2$ is the Kerr coefficient. Further, $z$ is the propagation distance, $L$ is the total length of the grating, $n_0$ is the average refractive index with modulation depth $\Delta n$, $\Lambda_0$ is the BG period at the input edge, and $C$ is the chirp. The wavenumber-detuning parameter δ (Figure 6), and the coupling between the forward and backward field, $\kappa = \pi \Delta n / \Lambda_0$, are functions of the propagation distance.

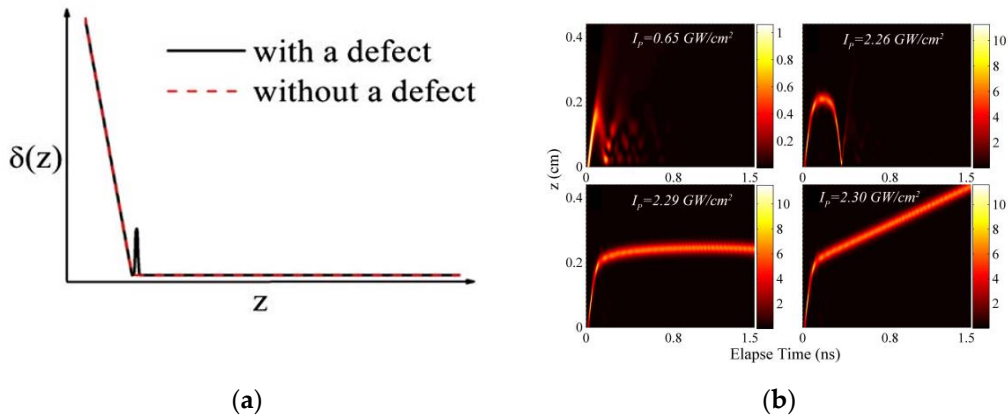

(**a**)          (**b**)

**Figure 6.** (**a**) The relation between wavenumber detuning and propagation distance in the concatenated BG, which does not include the local defect; (**b**) The evolution of the pulse with peak intensity equal to (d₁) 0.65 GW/cm², (d₂) 2.26 GW/cm², (d₃) 2.29 GW/cm², and (d₄) 2.30 GW/cm² [14].

The pulse-propagation properties are first investigated with a different input-pulse peak intensity (Figure 6). The analysis leads to the following conclusions.

(i) At low intensities, e.g., 0.65 GW/cm², the pulse is almost totally reflected by the Bragg structure due to the presence of the photonic bandgap (Figure 6bd₁). Since the pulse's dispersion is not compensated by the weak nonlinearity, pulse stretching is observed.

(ii) At higher intensities, an unstable standing light pulse trapped at the interface is generated. In particular, at $I_P$ = 2.26 GW/cm², the pulse will be reflected after stopping for a short time, as seen in panel (d₂) of Figure 6b. Slightly increasing $I_P$ to 2.29 GW/cm², in the range of 2.26 to 2.29 GW/cm², gives rise to a stopping time of ~1.3 ns for the pulse that is eventually reflected, as shown in panel (d₃).

(iii) A slow moving Bragg soliton can be observed, as shown in panel (d₄), for $I_P$ = 2.30 GW/cm². In this case, the velocity of the pulse is equal to 0.005 c. Further simulations demonstrate that the velocity of the moving soliton increases with a further increase of $I_P$.

Since the standing pulse is unstable, a defect is introduced between the chirped and uniform BG segments, as shown in Figure 6a. By introducing such a defect, the stability of pulse trapping can be much improved. Additionally, the intensity range of achieving a standing pulse can be made several times larger than without the defect (Figure 7).

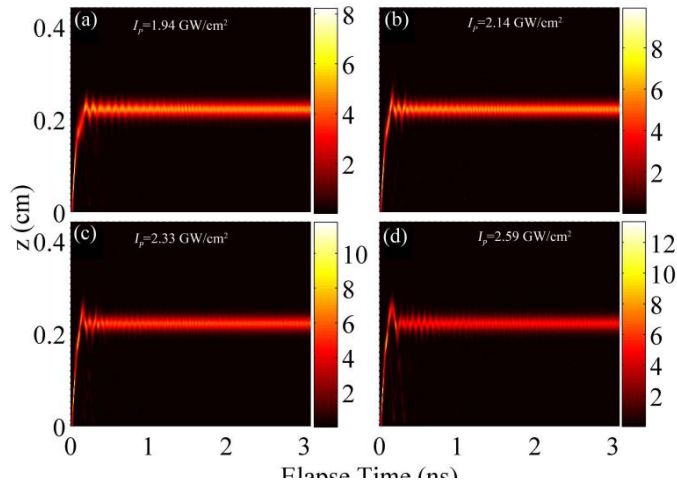

**Figure 7.** Simulations of the pulse propagation in the concatenated BG, with a defect located at the conjunction of the chirped and uniform segments, for different injected peak intensities: (**a**) 1.94 GW/cm$^2$; (**b**) 2.14 GW/cm$^2$; (**c**) 2.33 GW/cm$^2$; and (**d**) 2.59 GW/cm$^2$ [14].

To control the trapping position, a periodic set of defects is introduced into the uniform part of the BG structure, as shown in Figure 8, which shows $\delta_d(z)$ as a function of *z*. Here, ε is the strength of each defect and $d_w$ is its width [15].

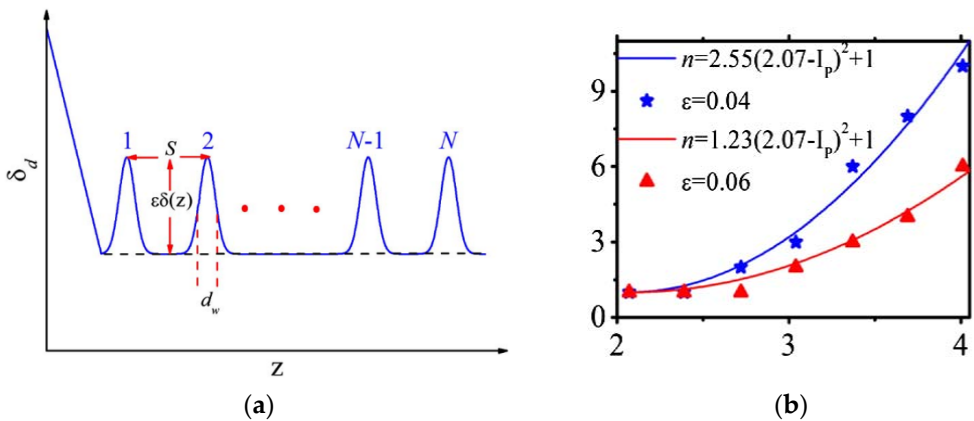

(**a**) (**b**)

**Figure 8.** (**a**) A schematic of the system, built of the linearly chirped BG segment (on the left-hand side) followed by the uniform grating with an inserted periodic array of local defects. The system can be described by parameters (*S*; $d_w$; ε), and the definition of *S*, $d_w$, and ε are shown in (**a**); (**b**) Relations between the trapping position and the initial pulse's intensity $I_P$ for (*S*; $d_w$; ε) = (0.132 cm; 50 μm; 0.04) and (0.132 cm; 50 μm; 0.06) (blue and red curves, respectively) [15].

The results of the simulations show that, at different input-pulse intensities, the pulse can be trapped at different defects. Such a defect array can also trap several pulses at different defects (Figure 9). In addition to that, the trapped pulses can interact with each other.

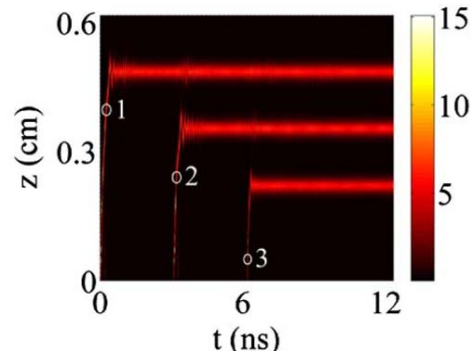

**Figure 9.** The simulation result of the trapping of three pulses in the BG structure. Here ($S$; $d_w$; $\varepsilon$) = (0.132 cm; 50 µm; 0.04). The peak intensities of the three pulses are 3.04, 2.78, and 2.07 GW/cm². Such pulses are launched into the gratings at $t$ = 0, $t$ = 3, and $t$ = 6 ns, respectively [15].

An all-optical femtosecond soliton diode can also be designed, using a tailored BG heterojunction. Highly nonreciprocal transmission, with an extinction ratio of up to 120, was produced by simulations of this setting (see Figure 10).

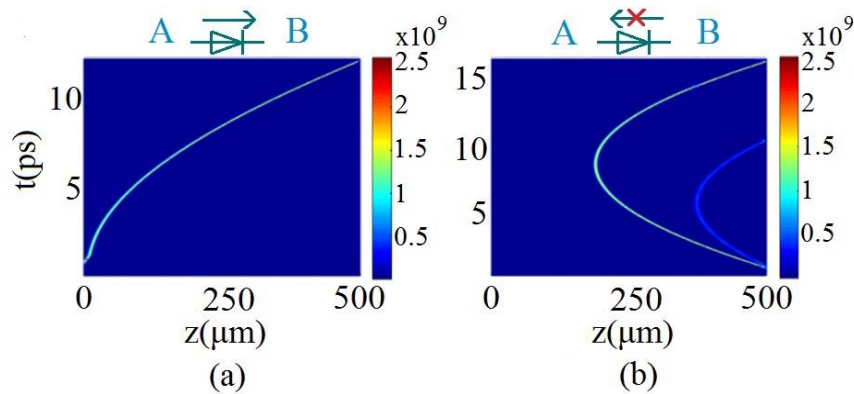

**Figure 10.** A typical example of the predicted femtosecond diode effect. The pulses with a width of 100 fs are injected from the left- (**a**) and right- (**b**) hand sides, respectively. The incident peak intensity of the pulse is $I$ = 0.35 GW/cm². (**a**) The femtosecond soliton can propagate from A to B; (**b**) the soliton bounces back when it is injected from B [16].

## 5. Ultrafast Pulse Modulations Based on Cholesteric Liquid Crystal (CLCs) Bragg Gratings

For modulating the optical signal in a compact photonic device, the following properties of the device materials are required: (i) Fast response to process a high speed optical signal; and (ii) Strong dispersion and nonlinearity to achieve a shorter dispersion and nonlinear length to decrease the required propagation length. Many works on the pulse modulation have been completed in fiber Bragg grating, and for which, the typical nonlinear coefficient is on the order of $10^{-16}$ cm²/W; in conjunction with a peak power of kilowatts [26], the resulting nonlinear effective length is on the millimeter scale [60]. In this section, we show that naturally occurring Bragg grating in a highly nonlinear CLCS can enable the same ultrafast (femtoseconds-picoseconds) pulse modulation operations in sub-mm interaction lengths.

After decades of studies, liquid crystals have emerged as highly versatile and nonlinear optical materials, due to their organic molecular constituents and unique liquid crystalline properties [61].

In liquid crystals, the underlying mechanisms generally fall into two classes, as depicted in Figure 11 for the nematic phase (which include the chiral nematic phase often called cholesteric liquid crystals (CLCs)): (i) Macroscopic crystalline responses including the director axis, reorientation by light fields or a light/DC field induced photorefractive effect, index/birefringence change by laser induced order parameter and/or temperature changes, and flow reorientation; these bulk effects generally respond in the millisecond to a sub-microsecond regime [62]; and (ii) Individual molecule's nonlinear polarization associated with single- and multiple photonic transitions within the molecular energy levels; these respond on the sub-picoseconds to a femtoseconds scale [63–67].

CLCs are formed by introducing achiral agent in standard nematic liquid crystals used in ubiquitous liquid crystal display devices. In CLC, the birefringent molecules self-assemble with the director (crystalline) axis arranged in a spiral, with a pitch on the order of the optical wavelength stretching from the UV to the infrared regime, as seen in Figure 12. As a result, CLC's possess not only the advantageous features of liquid crystals such as fabrication ease and low cost, and a very wide spectral dynamic range of operation [68,69], but also the 1-D photonic crystals' unique ability to enhance the nonlinear ultrafast all-optical responses of the CLC constituent molecules [mostly due to the nematic constituent], in addition to the dispersion effect at the photonic band-edges. A typical magnitude of the non-resonant nonlinear index coefficient $n_2$ is in the order of $10^{-14}$–$10^{-13}$ cm$^2$/W, but owing to the photonic crystal band-edge enhancement, the magnitude of the effective $n_2$ can be as large as $10^{-12}$–$10^{-11}$ cm$^2$/W. In comparison to other materials [70] used for ultrafast pulsed laser modulation applications, these $n_2$ values are orders of magnitude larger, and thus, one can envision a tremendous miniaturization possibility.

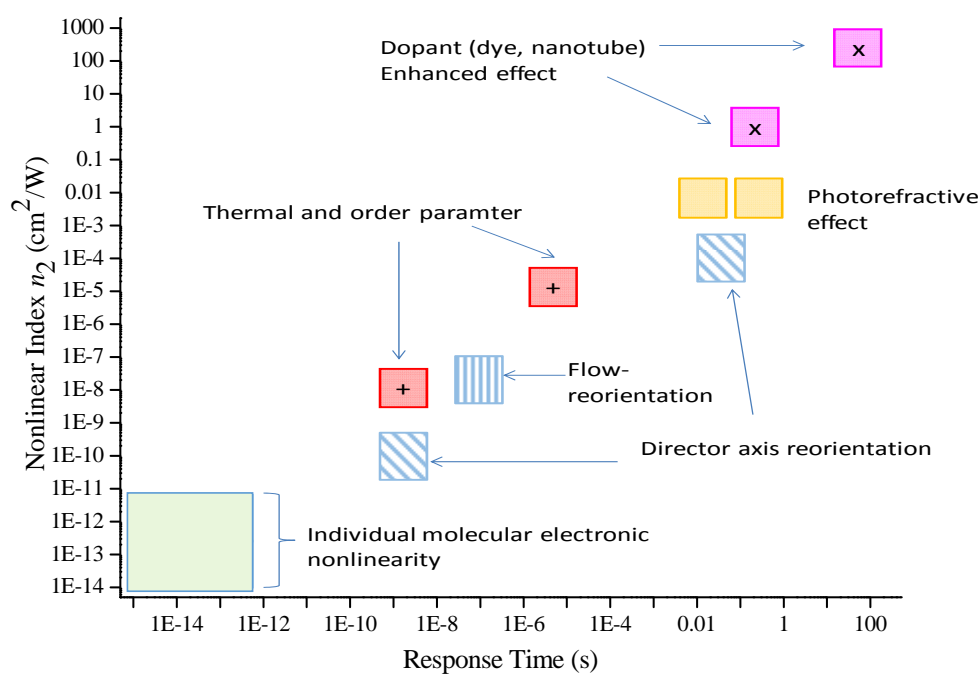

**Figure 11.** Observed optical nonlinearity in terms of the nonlinear index coefficients of nematic liquid crystals (including chiral nematics) for several mechanisms and the characteristic relaxation time constants.

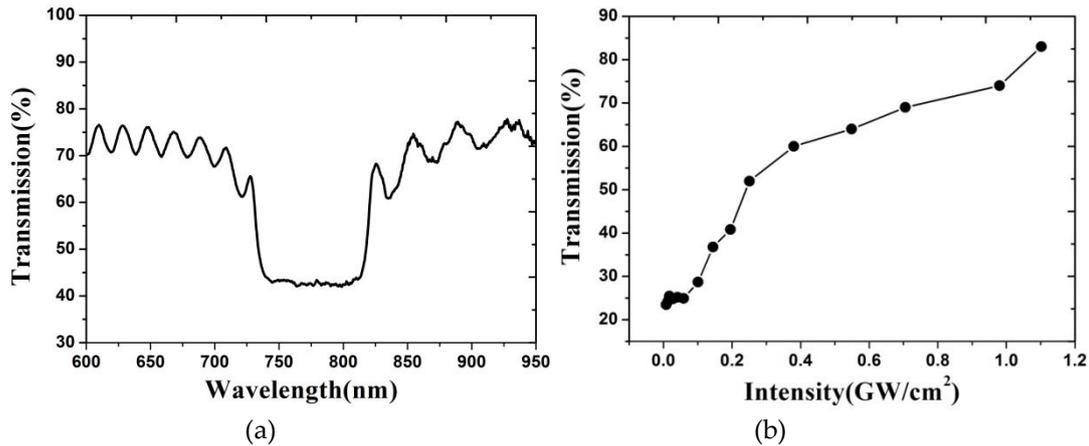

**Figure 12.** (**a**) Observed transmission spectrum of linearly polarized light through the CLC cell used in the experiment; (**b**) Observed dependence of the transmission of a left-handed circularly polarized laser pulse ($\lambda$ = 815 nm) on the peak intensity [17].

Depending on the electronic resonances of the nematic constituent used in synthesizing a CLC and the wavelength of the laser under study, the sign of $n_2$ can be positive [18] or negative [17]—a common feature found in the electronic nonlinearities of most materials [70]. As shown in Figure 12b for the nematic compound used in [18], the refractive index change induced by the laser causes the bandgap to shift towards the shorter wavelength region (i.e., blue-shit), and consequently gives rise to an increasing intensity dependence since the laser wavelength is located at the long wavelength edge of the CLC (Figure 12b). In Reference [17], the CLC sample with self-defocusing nonlinear properties is used to compress the femtosecond pulse (Figure 13). The nonlinear coefficient of the CLC used in that study was measured to be about −10$^{-11}$ cm$^2$/W using a similar intensity dependent transmission measurement. This is four orders of magnitude higher than silica [12]; as a result, the required the thickness of the CLC sample is merely 6 microns in order to compress a 100 fs laser pulse to about 50 fs [17]. If the constituent nematic molecules possess a lower optical nonlinearity, thicker CLC cells of several 100's microns are required[18], which are nevertheless thin/short compared to other materials used for ultrafast laser pulse modulations.

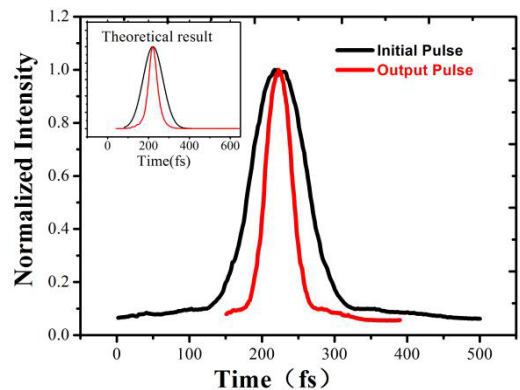

**Figure 13.** Observed direct pulse compression effect on an initial transform-limited 100 fs pulse (black line) to a 48 fs output pulse (red line); the inset figure corresponds to the simulation results using the measured experimental parameters [17].

The pulse propagation can be simulated by the nonlinear coupld-mode theory presented in part 4. In [17,18], it is shown that the simulations for the pulse width and the spectra are in good agreement with the experimental results, as seen in Figures 13 and 14. The compression ability for the pulse with different widths is determined by the propagation length, since the differences of dispersion for different widths lead to a difference in the dispersion length. Another work reported in Reference [18] shows that by using a CLC sample with a 500 μm cell gap, the sub-picosecond pulse with 800 fs can be compressed to 286 fs. The corresponding spectra broadening can also be observed (Figure 15). This result clearly indicates that by cascading the CLC sample with a different cell gap, the pulse can be compressed in a wide temporal range.

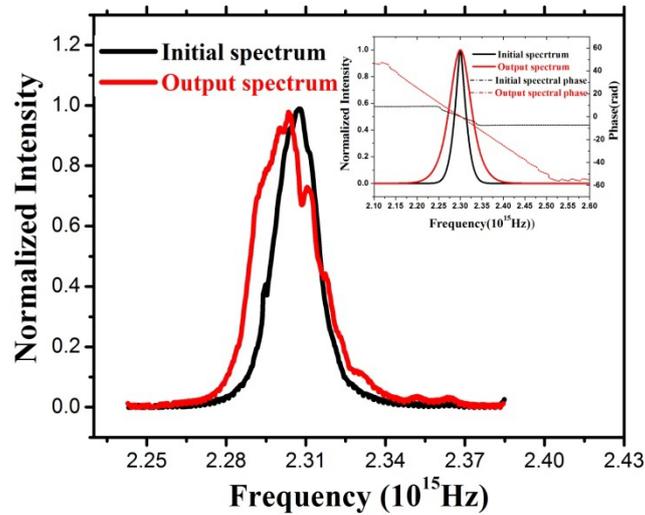

**Figure 14.** Observed spectral broadening effect due to the pulse compression for an input pulse peak intensity of 1.04 GW/cm²; the inset depict simulation results for the spectrum (solid lines) and spectral phase (dashed lines) [17].

Another possible pulse-modulation device is based on two tandem CLC cells that are tailor-made so that their respective blue/red photonic band-gap edges match the operating laser wavelength, and to utilize the opposite linear dispersions (i.e., without involving the optical nonlinearity) from these band edges to impart an opposite chirping effect on these pulses. As a result, an incident 100 femtoseconds laser pulse can be stretched to nearly 2 picoseconds by the first CLC cell, and then recompressed to the original pulse length by the second CLC cell, as seen in Figure 16. Such pulse stretching and recompressing operations are commonly employed in chirped pulse amplification (CPA) systems [71] that require an intermediate pulse stretching process to prevent amplification saturation. In this case, the entire stretch-recompress operation can be done all-optically with CLC cells measuring <1 mm in interaction length, contrasting greatly with the bulky optics employed in conventional CPA systems.

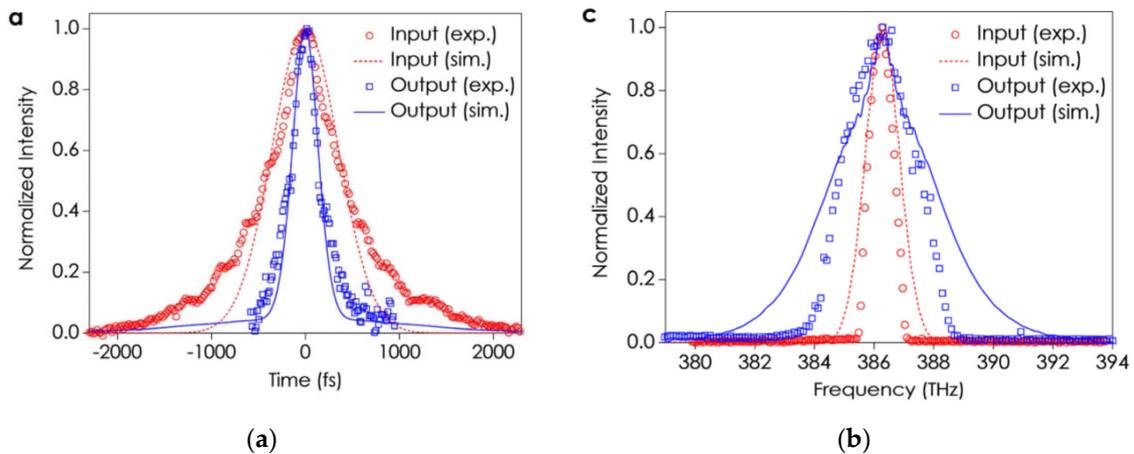

**Figure 15.** (**a**) Observed direct pulse compression effect on an initial transform-limited 847 fs pulse (open circles) to a 286 fs output pulse (open squares); the inset figure corresponds to the simulation results using the measured experimental parameters; (**b**) Observed and simulated spectra for the input and output (compressed) pulses [18].

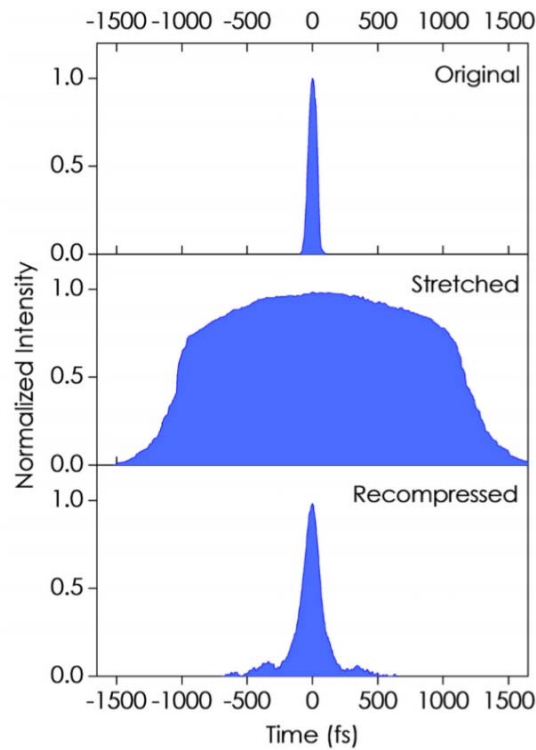

**Figure 16.** Results of pulse modulations by two tandem 550 μm thick CLC cells. Observed pulse shape for the input (upper trace), output after traversing the first cell (middle), and output after traversing the second cell (bottom trace). Initial pulse width: 100 fs laser pulse; wavelength: 780 nm [18].

## 6. Conclusion

We have discussed several methods to realize an ultrafast optical signal process in 1D Bragg structures, including resonantly absorbing Bragg reflectors (RABR), chirped BGs (Bragg gratings), and transparent but highly nonlinear CLCs (cholesteric liquid crystals) that exhibit properties of 1D photonic crystals. The result shows that an ultrafast pulse can be buffered and subsequently released in the RABR Additionally, based on the reduction of the light speed, a nonlinear frequency conversion can be achieved over a short propagation distance. Using a linearly chirped BG concatenated with a uniform one, we have demonstrated the possibility of achieving the efficient slow-down of light with the right mix of forward and backward propagating fields. Such chirped gratings can also function as all-optical photodiodes. Finally, we have discussed a unique class of self-assembled BGs based on CLCs. They feature extraordinarily large and ultrafast-responding optical nonlinearities, which enable direct compressions, as well as the stretching and recompressions of femtoseconds laser pulses over very short (sub-mm) propagation lengths.

**Acknowledgments:** This work is supported by the Chinese National Natural Science Foundation (61505265, 11374067, 11534017). Yikun Liu is also supported by the Fundamental Research Funds for the Central Universities. Iam Choon Khoo's travel and local support is provided by the William E. Leonhard Professorship of PSU.

**Author Contributions:** J.Z. coordinated the work on the paper. Y.L. collected the data and wrote the draft. S.F., B.A.M., and I.C.K. took part in drafting the paper and contributed to the critical analysis of the data.

**Conflicts of Interest:** The authors declare no conflict of interest.